\documentclass[11pt]{revtex4-1}

\usepackage{graphicx}

\begin{document}

\title{ A Dynamical System Analysis of Three Fluid cosmological Model}

\author{Nilanjana Mahata\footnote {nmahata@math.jdvu.ac.in}}
\author{Subenoy Chakraborty\footnote {schakraborty@math.jdvu.ac.in}}
\affiliation{Department of Mathematics, Jadavpur University, Kolkata-700032, West Bengal, India.}



\begin{abstract}
In Friedman-Robertson-Walker  flat spacetime, we consider a three fluid cosmological model which contains dark matter, dark energy and baryonic matter in the form of perfect fluid  with a barotropic equation of state. Dark matter is taken in form of dust and dark energy is described by a scalar field with a potential $ V(\phi) $.
Einstein's field equations are reduced to an autonomous dynamical system by suitable redefinition of basic variables. Considering exponential potential for the scalar field, critical points are obtained for the autonomous system. Finally stability of the critical points and cosmological implications are analyzed.\\

\end{abstract}

\maketitle



\section{Introduction}
A large number of recent observational  data, including Type Ia supernovae \cite{RefA1, RefA2}, Large Scale Structure \cite{RefA3, RefA4} and Cosmic Microwave Background anisotropies \cite{RefA5,RefA6}, Baryon Accoustic Oscillations \cite{RefA7}  have provided  evidences for a spatially flat universe which suffered two accelerated phases
- an early acceleration phase (inflation), which occurred prior to the radiation dominated era and a
recently initiated present time accelerated expansion. This late time  accelerated expansion is attributed to  a mysterious exotic matter with large negative pressure called  \textit{dark energy}. The nature of dark energy is unknown but the  feature of dark energy is that it remains unclustered at all scales whereas gravitational
clustering of baryons and nonbaryonic cold dark matter can be seen. The combined analysis of the different
cosmological observations suggests that the universe consists of about 70 \%   dark energy, 25 \% dark matter and 5\%   baryons and  radiation. As the origin of dark energy is unknown, so several candidates have been proposed to describe it. The simplest choice for dark energy  is the cosmological constant \cite{RefA8,RefA9,RefA10}. This so-called $\Lambda $CDM model is able to explain most of the current observational data. But it suffers from two problems - \textit{coincidence problem} and \textit{fine tuning problem}.  The lack of a reasonable explanation for the cosmological constant problems has led researchers to explore other routes to explain the observations. So dynamical dark energy models have been proposed as alternatives in the literature. A wide class of scalar field models have been introduced to model dark energy. Scalar field models  including \textit{quintessence} \cite{RefA11,RefA12},  \textit{K-essence}\cite{RefA13, RefA14}, tachyon etc have attracted lots of attention \cite{RefA15,RefA16,RefA17,RefA18,RefA19}.\\
In this paper, we consider a  FRW model of the Universe which contains three types of matter - dark matter, dark energy and baryonic matter \cite{RefA20}. The dark matter  is in form of dust.  The dark energy is described by a scalar field and the baryons are described by a perfect fluid. As a  scalar field can be considered to be a perfect fluid  in effect, so  the matter in the universe  is considered in terms of two perfect fluids and a dust. All three fluids are assumed to be self-interacting and minimally coupled to gravity.\\
We will study the model qualitatively and will check for viable cosmological solution considering cosmological constraints and observational data. From Einstein's field equations along with Klein Gordan equation we form an autonomous system. In order to study  the dynamical character of the system, critical points are obtained and  corresponding cosmological models are analyzed. Feasible  cosmological solutions should depict our present universe as global attractor i.e all the possible initial conditions lead to the observed percentages of dark energy and dark matter, and  once reached, they remain fixed forever.  For this reason we will focus on the stability of critical points i.e cosmological models which are attractors.  For theoretical details of dynamical system method we would suggest to go through the books \cite{RefA22, RefA23, RefA21}.\\
 The paper is organized as follows: next section describes the basic equations and physical background of the model. Section III describes formation of autonomous system. Phase space analysis for both 3D and 2D autonomous system are discussed in detail in section IV. Summary is written in section V.

\section{Basic Equations}
The standard cosmology suggests that our universe went through the radiation dominated era, followed by matter dominated phase  and then according to recent observations the universe is going through  an accelerated expansion. For consistence any theoretical model should coincide with the history of the universe. The homogeneous and isotropic flat FRW space time is chosen as the model of the universe. The universe  is assumed to be filled up with non interacting \textit{dark matter}, \textit{dark energy }and \textit{baryon}. Dark matter is considered  in the form of dust having  energy density $\rho_{m}$ and  dark energy  is driven by a scalar field    $ \phi $  with potential $ V(\phi)$ whose energy density  and pressure is given by
\begin{equation}\label{eqn(1)}
 \rho_{d} = \frac{1}{2}\epsilon \dot{\phi}^{2} + V(\phi) ~~~and  ~~~  p_{d} = \frac{1}{2}\epsilon \dot{\phi}^{2} - V(\phi)
\end{equation}
respectively. We will have quintessence ( real $\phi$) if $ \epsilon= 1 $ or phantom field (imaginary $\phi$) if $ \epsilon = -1 $. We have chosen $\epsilon =1$ here.

The \textit{'baryonic  matter'} is described by a perfect fluid with linear equation of state $ p_{b}= (\nu - 1) \rho_{b} $ where $\rho_{b} , p_{b} $  are density and pressure of the fluid and satisfies $ \frac{2}{3} < \nu \leq 2 $ where $ \nu $ is the adiabatic index of the fluid. In particular $\nu = 1 $ and $\nu = \frac{4}{3}$ correspond to dust matter and radiation respectively.
 All three fluids interact minimally.\\
The Einstein field equations can be written as (here $ k = 8\pi G $ is gravitational constant, c = 1 )
\begin{equation}\label{eqn(2)}
 3H^{2} = k(\rho_{m} + \rho_{d} + \rho_{b})
\end{equation}
\begin{equation}\label{eqn(3)}
   2\dot{H}  = - \frac{k}{2}( \rho_{m} + \rho_{b} + p_{b} +  \dot{\phi}^{2})
\end{equation}
where H is the Hubble parameter.\\
Klein-Gordan equation for the scalar field is
\begin{equation}\label{eqn(4)}
   \ddot{\phi}  + 3H \dot{\phi} +  \frac{dV}{d\phi} = 0
\end{equation}

The energy conservation relations take the form
\begin{eqnarray}
\dot{\rho_{m}}+ 3H\rho_{m} &=& 0 \label{eqn(5)}\\
\dot{\rho_{d}}+ 3H(\rho_{d} + p_{d})& =& 0  \label{eqn(6)}\\
\dot{\rho_{b}}+ 3H(\rho_{b} + p_{b}) &= &0   \label{eqn(7)}
\end{eqnarray}

From (\ref{eqn(5)}) we get, $ \rho_{m} = \frac{A}{a^{3}} $, A is an integration constant, 'a(t)' is scale factor of the universe.
Equation (\ref{eqn(7)}) implies that $ \rho_{b} = Ca^{-3\nu}$ and $ p_{b} = (\nu -1) Ca^{-3\nu}$, where  $ \frac{2}{3} <  \nu  \leq 2, \nu \neq 1 $.\\
Now  from  (\ref{eqn(2)}), (\ref{eqn(3)}) and (\ref{eqn(4)}), we obtain the system of equations as
\begin{equation}\label{eqn(8)}
 3H^{2} = k(\frac{A}{a^{3}} + \frac{1}{2}  \dot{\phi}^{2} +  V(\phi)   +  Ca^{-3\nu})
\end{equation}
\begin{equation}\label{eqn(9)}
   2\dot{H}  = - \frac{k}{2}( Aa^{-3}+ \nu Ca^{-3\nu} +  \dot{\phi}^{2})
\end{equation}
\begin{equation}\label{eqn(10)}
   \ddot{\phi}  + 3H \dot{\phi} +  \frac{dV}{d\phi} = 0
\end{equation}

\section{ Formation  of Autonomous system }
The above evolution equations are highly non-linear, so we
 will study our cosmological model qualitatively. At first we will formulate an autonomous system from (\ref{eqn(8)})- (\ref{eqn(10)}). We will discuss the properties and cosmological consequences based on the existence, stability of critical points and value of the cosmological parameters. We employ the dimensionless variables \cite{RefA19, RefA24},
\begin{equation}\label{eqn(11)}
 x = \sqrt{\frac{k }{6}}\frac{\dot{\phi}}{H} ~~~ and ~~~ y = \sqrt{\frac{k}{3}}\frac{\sqrt{V(\phi)}}{H}
\end{equation}

It is conventional to write fractional energy densities as
\begin{equation}\label{eqn(12)}
   \Omega_{m} = \frac{kAa^{-3}}{3H^{2}} ~~~~and~~~~  \Omega_{b} = \frac{kCa^{-3\nu}}{3H^{2}}
\end{equation}

Now the  Friedmann equation (\ref{eqn(8)}) can be rewritten as
\begin{equation}\label{eqn(13)}
   \Omega_{m} +  \Omega_{b} + x^{2} + y ^{2} = 1
\end{equation}

From (\ref{eqn(9)}), we have
\begin{equation}\label{eqn(14)}
\dot{H}= -{3H^{2}}(x^{2} + \frac{ \Omega_{m}}{2} + \frac{\nu \Omega_{b}}{2} )
\end{equation}
Now, $ 0 \leq \Omega_{b} \leq 1 \Rightarrow $
\begin{equation}
0 \leq \Omega_{m} + x^{2} + y ^{2} \leq 1 \label{eqn(15)}
\end{equation}

The evolution of the dynamical system is described by  the autonomous system
\begin{eqnarray}\label{eqn16}
  \frac{dx}{dN}& = &3x(x^{2} - 1 + \frac{ \Omega_{m}}{2}+ \frac{\nu \Omega_{b}}{2})- \sqrt{\frac{3  }{2k}} \frac{ V'(\phi)}{V(\phi)}y^{2}\\
\nonumber
 \frac{dy}{dN}& = &y[3(x^{2} + \frac{\nu \Omega_{b}}{2} + \frac{ \Omega_{m}}{2}) + \sqrt{\frac{3}{  2k}} \frac{ V'(\phi)}{V(\phi)}x]\\
 \nonumber
\frac{d\Omega_{m}}{dN}& =& -3 \Omega_{m}[ 1 - \Omega_{m} -2x^{2} - \nu \Omega_{b}]\\
\nonumber
\end{eqnarray}
where  $ N \equiv ln~ a $.\\
The effective equation of state for this three fluid model has the expression
\begin{equation}\label{eqn17}
\omega_{eff} = \frac{ p_{\phi} + p_{b}}{\rho_{m} + \rho_{d} + \rho_{b}} = -1 + 2x^{2}+ \nu( 1 - \Omega_{m} - x^{2} - y^{2})+ \Omega_{m}
\end{equation}
For cosmic acceleration,   $ \omega_{eff} < -\frac{1}{3}$  is required.\\
The equation state for dark energy is given by
\begin{equation}\label{eqn18}
   \emph{} \omega_{\phi} = \frac{p_{\phi}}{\rho_{\phi}} = \frac{x^{2}-y^{2}}{x^{2}+ y^{2}}
\end{equation}
and the density parameter for the scalar field is
\begin{equation}\label{eqn19}
   \Omega_{\phi} = \frac{k\rho_{\phi}}{3H^{2}}= x^{2}+ y^{2}
\end{equation}
It should be noted that the physical region in the phase plane is constrained by the requirement that the energy density be non-negative i.e $\Omega_{m}\geq 0$ and $ \Omega_{b} \geq 0$. So (\ref{eqn(15)}) restricts the dependent variables x and y to be on the circular cylinder $ x^{2}+ y^{2}\leq 1$.
 The equality sign indicates that there is no longer any dark matter.

\section{Phase space analysis  and critical points}
\subsection{3D autonomous system}

We assume  $ \frac{ V'(\phi)}{V(\phi)} $ =  constant i.e potential of the scalar field is exponential, then we can write
$\sqrt{\frac{3}{  2k}} \frac{ V'(\phi)}{V(\phi)} $ = constant = $\alpha $ (say). 

We rewrite the autonomous system as
\begin{eqnarray}\label{eqn20}
\frac{dx}{dN}& = &3x[x^{2} - 1 + \frac{\Omega_{m}}{2}+ \frac{\nu}{2}(1 - \Omega_{m} - x^{2} - y^{2})]- \alpha y^{2}\\
 \frac{dy}{dN}& = &y[3x^{2} + 3 \frac{\nu }{2}(1 - \Omega_{m} - x^{2} - y^{2}) + 3\frac{ \Omega_{m}}{2} +   \alpha x]\\
\frac{d\Omega_{m}}{dN}& = &-3 \Omega_{m}[ 1 - \Omega_{m} -2x^{2} - \nu(1 - \Omega_{m} - x^{2} - y^{2}) ]\label{eqn22}
\end{eqnarray}

In order to obtain the critical points of the system (\ref{eqn20})-(\ref{eqn22}), we set $\frac{dx}{dN}=0$, $\frac{dy}{dN}= 0 $
and $\frac{d\Omega_{m}}{dN}= 0$. An attractor is one of the stable critical points of the system. Of course, the critical points need to satisfy (\ref{eqn(15)}) for validity. Clearly, (0, 0, 0) is a critical point. Other critical points are ( 0, 0, 1 ), ( 1, 0, 0 ), ( -1, 0, 0 ),(  0, 0, $\Omega_{m}$) ~if $\nu = 1 $ and ( $\frac{3}{2\alpha}(\nu - 1), \pm \frac{3}{2\alpha}(1- \nu ), 1-\frac{9}{2}\frac{(1-\nu)^2}{\alpha^2}$).\\
 Considering linear perturbations about the critical point ($ x_{c},y_{c},\Omega_{mc} $) into the dynamical system equations, the  linearized perturbation matrix takes the form
 \[
   A =
  \left[ {\begin{array}{ccc}
 9x_{c}^2(1-  \frac{\nu}{2})-3(1-\frac{\Omega_{mc}}{2})&  -3\nu y_{c}x_{c}- 2\alpha y_{c} & \frac{3x_{c}}{2}(1-\nu) \\
 +3(1-\Omega_{mc})\frac{\nu}{2}&& \\
 &&\\
   3y_{c}x_{c}(2-\nu) + \alpha y_{c} & 3x_{c}^2(1- \frac{\nu}{2})+\frac{3\nu}{2}+ \frac{3\Omega_{mc}}{2}(1- \nu)  & \frac{3x_{c}}{2}(1-\nu) \\
   & + \alpha x_{c}  -\frac{9}{2}\nu y_{c}^2&\\
   &&\\
    -6x_{c}\Omega_{mc}(\nu -2) & -6\nu y_{c}\Omega_{mc} & 6\Omega_{mc}(1-\nu)+3x_{c}^2(2-\nu)\\
    &&-3\nu y_{c}^2-3(1-\nu)\\
  \end{array} } \right]
\]
Hyperbolic critical points of the system, related cosmological parameters, eigen values are given in table I.

\begin{table}
 \caption{\emph{Equilibrium points, related  parameters and Eigen values } }
 \begin{tabular}{|c|c|c|c|c|c|c|p{1.8 in}|}
 \hline
   Equilibrium point & x & y  & $  \Omega_{m} $ & $ \Omega_{\phi}$ & $ \Omega_{b}$ & $ \omega_{eff}$  & Eigen Values \\\hline
 $ C_{1}$  & 0 & 0 & 0 & 0 & 1 & $  -1+ \nu $ & $ -3(1- \nu), \frac{3\nu}{2},3(\frac{\nu}{2}-1) $ \\\hline
 $ C_{2}$ & 1 & 0 & 0 & 1 & 0 &1 &  $ 6(1-\frac{\nu}{2}) $ , $3+\alpha $,3 \\\hline
 $ C_{3}$ & -1 & 0 & 0 & 1 &0 &1 &  $ 6(1-\frac{\nu}{2}) $ , $3 - \alpha $,3 \\\hline
 $ C_{4}$ & 0 & 0 & 1 & 0 &0 &0 &  $ -\frac{3}{2},\frac{3}{2}, 3(1 - \nu) $ \\\hline
 $ C_{5}$  &$ \frac{3}{2\alpha}(\nu-1)$ & $\pm \frac{3}{2\alpha}(1-\nu)$ & $ < 1-\frac{9}{2}\frac{(1-\nu)^2}{\alpha^2}$ &  $ \frac{9}{2}\frac{(1-\nu)^2}{\alpha^2}$ & $>0$& 0  &   in table II  \\\hline

 \end{tabular}
 \end{table}

$ C_{1}$ represents  non acceleating universe completely dominated by baryonic matter. 
 This equilibrium point is unstable because one eigen value is positive. $ C_{2}$ and  $ C_{3}$ represents universe dominated by dark energy and these two points are   unstable, one eigen value being positive. Late time acceleration is denied for these points. The stability of dark energy dominated phase will determine the fate of the universe.  The critical point $ C_{4}$ is dominated by dark matter  describing  a non accelerating universe. It is unstable in nature. From history of the universe, we expect that the matter dominated era should be unstable, otherwise it would not enter dark energy dominated phase.  There should  exist unstable matter dominated phase along with stable dark energy dominated phase. From the table I, we see that no critical point matches with present day observation. There is another critical point $(0,0,\Omega_{m}),~ if ~ \nu = 1 $ which is non hyperbolic in nature as two eigen values are zero. Hence, we can not investigate the local stability of the system at this point and it does not have any stable manifold by center manifold theorem. This model represents completely dark matter dominated phase without late time acceleration.

For different values of $\nu$ and $\alpha$, $C_{5}$ will denote different critical points. These points are
listed in table II with their eigen values. $C_{5a}$, $C_{5b}$ (in table II) denotes same critical point but for different $\nu $ and $ \alpha $. We can see that $C_{5a}$, $C_{5b}$, $C_{5e}$ and $C_{5h}$ have all the eigen values negative. So we can say that they may represent stable phase. Except $ C_{5h}$, the corresponding stable universe is largely dominated by dark energy, and dark matter density is very close to presently available observed value, but for $ C_{5h}$, the universe is dominated by dark matter.  The points do not explain late time acceleration although. Rest of the critical points  are unstable points.  $C_{5c}$, $C_{5d}$ and $C_{5f}$ are largely dominated by dark matter whereas $C_{5g}$ is dominated by dark energy. $C_{5c}$,$C_{5d}$ and $C_{5f}$  represent contracting universe as $y < 0 $ for these points.
None of these points describes viable model.

\begin{table}
 \caption{\emph{ $C_{5}$ for different  $\nu$ and $\alpha$, related  parameters and Eigen values } }
 \begin{tabular}{|c|c|c|c|c|c|c|c|c|p{1.8 in}|}
 \hline
   Equilibrium point& $\alpha$ & $\nu$ & x & y  & $  \Omega_{m} $ & $ \Omega_{\phi}$ &$  \Omega_{b} $& $ \omega_{eff}$  & Eigen Values \\\hline
 $ C_{5a}$ & -2.028 & 1.8  & -0.5916 & 0.5916 & .3 & .699 &.001& 0 & $ -0.0096 \pm 2.0543i, -0.2356 $ \\\hline
  $ C_{5b}$ & -1.52 & 1.6  & -0.5916 & 0.5916 & .3 & .699 & .001&0 & $ -0.027 \pm 1.8655i, -0.0055$ \\\hline
 $C_{5c}$ & 2 & 1.3  & .225 & -0.225 & .898 & .101 &.001 &0 & $ -0.9939 \pm .4031i, 1.6365$ \\\hline
$ C_{5d}$ & 1.3 & 1.2  & 0.375 & -0.375 & .718 & .281 &.001 &0 & $ -0.7040 \pm 0.7392i, 1.2323$ \\\hline

$ C_{5e}$ & -0.5 & 0.8  & 0.57 & 0.57 & 0.35 & .649 &.001& 0 & $ -0.527 \pm 0.594i, -0.3735 $ \\\hline
  $ C_{5f}$ & -1.5 & 0.9 & 0.1& -0.1 & .98 & .02 & 0 &0 & $ -1.446 , 1.304, -0.3055$ \\\hline
  $ C_{5g}$ & -0.5 & 1.2  & -0.6 & 0.6 & 0.28 & .72 &0 &0 & $ 0.1174 \pm 1.4189i, 0.1121 $ \\\hline
$ C_{5h}$ & 1.5 & 0.75 & -0.25 & 0.25 & .87 & 0.125 &.005 &0 & $ -0.828 \pm 0.18i, -1.21$ \\\hline

 \end{tabular}
 \end{table}

\subsection{2D Autonomous system}

The above 3D system i.e (\ref{eqn20})-(\ref{eqn22}) can be reduced to a 2D autonomous system if we do not consider time evolution of dark matter.  Recent observations predict that $\Omega_{m}$  is  approximately 0.30, and considering some bounds  we take  the  value of $\Omega_{m} $ within the range 0.24 - 0.34.

Then the system becomes
\begin{eqnarray}
\frac{dx}{dN}& = &3x[x^{2} - 1 + \frac{\Omega_{m}}{2}+ \frac{\nu}{2}(1 - \Omega_{m} - x^{2} - y^{2})]- \alpha y^{2}\label{eqn(24)}\\
 \frac{dy}{dN}& = &y[3x^{2} + 3 \frac{\nu }{2}(1 - \Omega_{m} - x^{2} - y^{2}) + 3\frac{ \Omega_{m}}{2} +   \alpha x]\label{eqn25}\\
 \nonumber
\end{eqnarray}

where $\Omega_{m}$ is constant lying within $ 0.24 < \Omega_{m} < 0.34 $.

\begin{table}
 \caption{ Critical points for $  \Omega_{\phi}=.32 $, $ \nu= 1.9$ and $ \alpha= 1.7$, related  parameters  }
 \begin{tabular}{|c|c|c|c|c|c|}
 \hline
   Equilibrium point &  x & y  &   $  \Omega_{\phi} $ & $  \omega_{eff} $ &  Nature \\\hline
 $ C_{2a} $  &  0  &  0 &  0 & 0.58 &  saddle  \\\hline
 $ C_{2b} $  & -0.44 & 0.76 & 0.77  & -0.29 & stable node \\\hline
 $ C_{2c}$  & -0.44 & -0.76 & 0.77 & -0.29 & stable node  \\\hline
 $ C_{2d}$ & 1.58  & 0 & 2.56  & 4.31 & unstable node \\\hline
$ C_{2e}$ & -1.48  & 0 & 2.19  & 4.31 & unstable  \\\hline
\end{tabular}
\end{table}

\begin{table}
 \caption{ Critical points for different $  \Omega_{m}= .3 $, $ \nu= 1.8$ and $ \alpha= -2 $, related  parameters  }
 \begin{tabular}{|c|c|c|c|c|c|}
 \hline
   Equilibrium point &  x & y  &   $  \Omega_{\phi} $ & $  \omega_{eff} $ &  Nature \\\hline
 $ C_{2a} $  &  0  &  0 &  0 & 0.56 &  saddle  \\\hline
 $ C_{3b} $  & .52 & .70 & 0.76  & -0.16  & stable node \\\hline
 $ C_{3c}$  & .52 & -.70 &0.76 & -0.16 & stable node  \\\hline

\end{tabular}
\end{table}

\begin{table}
 \caption{ Critical points for  $\Omega_{m} = 0.2 $, $ \nu= 1.6 $ and $ \alpha= 1.1 $, related  parameters  }
 \begin{tabular}{|c|c|c|c|c|c|c|}
 \hline
   Equilibrium point &  x & y  &   $  \Omega_{\phi} $ & $\Omega_{b}$ &$  \omega_{eff} $ &  Nature \\\hline
 $ C_{2a} $  &  0  &  0 &  0 & .8 & 0.48 &  saddle  \\\hline
 $ C_{4b} $  & -0.33 & 0.89 & 0.90 & .001 & -0.58  & stable node  \\\hline
 $ C_{4c}$  & -0.33 & -0.89 & 0.90 & .001& -0.58 & stable node  \\\hline

\end{tabular}
\end{table}

\begin{table}
 \caption{ Critical points for  $\Omega_{m} = 0.22$, $ \nu= 1.2$ and $ \alpha= 1.1$, related  parameters  }
 \begin{tabular}{|c|c|c|c|c|c|c|}
 \hline
   Equilibrium point &  x & y  &   $  \Omega_{\phi} $ &  $\Omega_{b}$ &$  \omega_{eff} $ &  Nature \\\hline
 $ C_{2a} $  &  0  &  0 &  0  & .78 & 0.156 &  saddle   \\\hline
 $ C_{5b} $  & -0.32 & 0.84 & 0.82  & .001 & -.57 & stable node  \\\hline
 $ C_{5c}$  & -0.32 & -0.84 & 0.82 & .001 & -.57 & stable node  \\\hline

\end{tabular}
\end{table}

\begin{table}
 \caption{ Critical points for  $\Omega_{m} = 0.24 $, $ \nu= 1.3 $ and $ \alpha= 3 $, related  parameters  }
 \begin{tabular}{|c|c|c|c|c|c|c|}
 \hline
   Equilibrium point &  x & y  &   $  \Omega_{\phi} $ &  $\Omega_{b}$ &$  \omega_{eff} $ &  Nature \\\hline
 $ C_{2a} $  &  0  &  0 &  0 &.76 & 0.228 & saddle   \\\hline
 $ C_{6b} $  & -0.58 & 0.47 & 0.56  & .20 & 0.17 & stable node  \\\hline
 $ C_{6c}$  & -0.58 & -0.47 & 0.56 & .20 & 0.17& stable node  \\\hline

\end{tabular}
\end{table}

\begin{figure}
\begin{minipage}{.45\textwidth}
 \includegraphics[width = 1.000\linewidth]{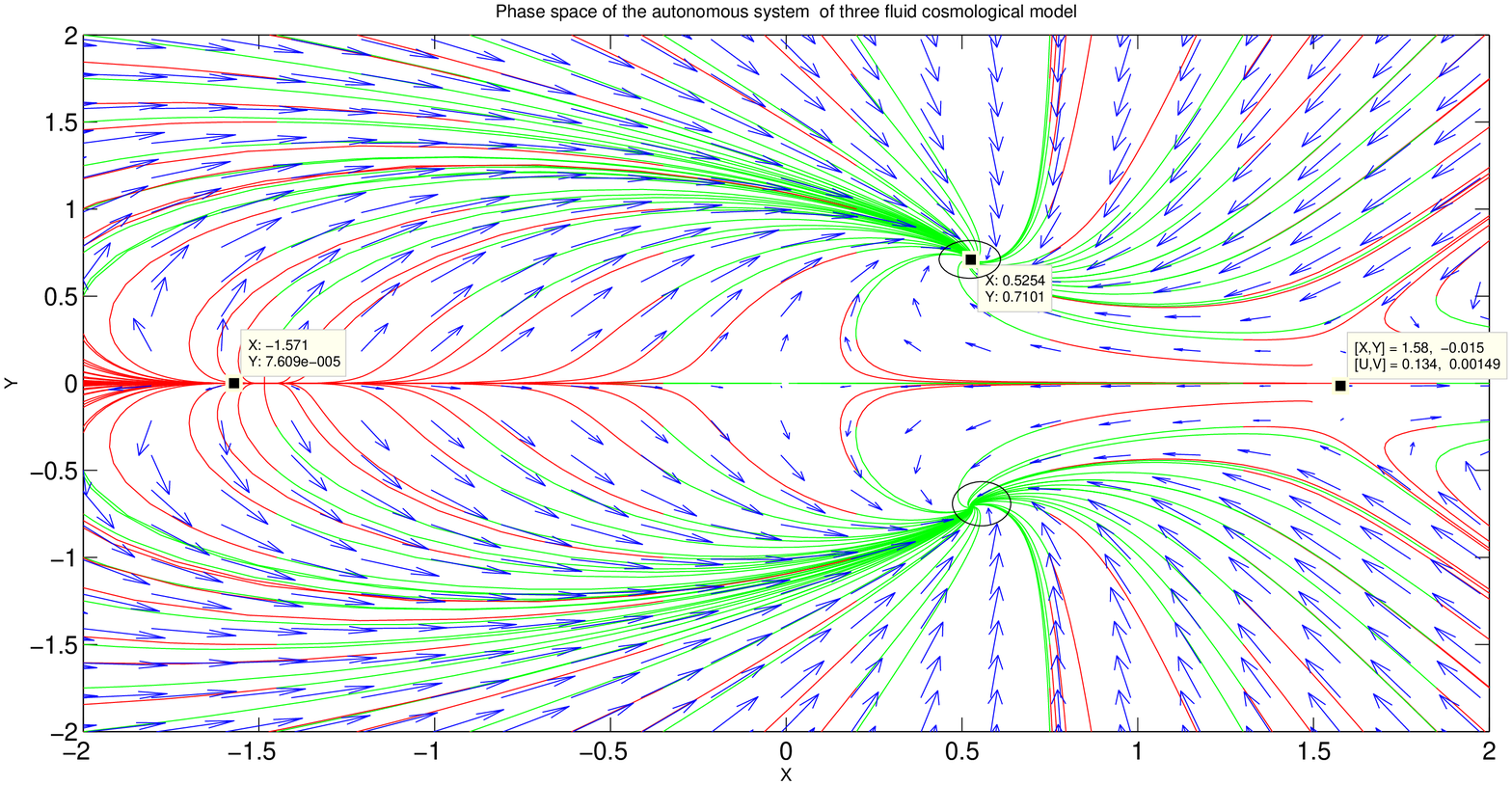}
\caption{\label{}Phase portrait for the system, related table III  }
 \end{minipage}
 \begin{minipage}{.45\textwidth}
 \includegraphics[width = 1.0\linewidth]{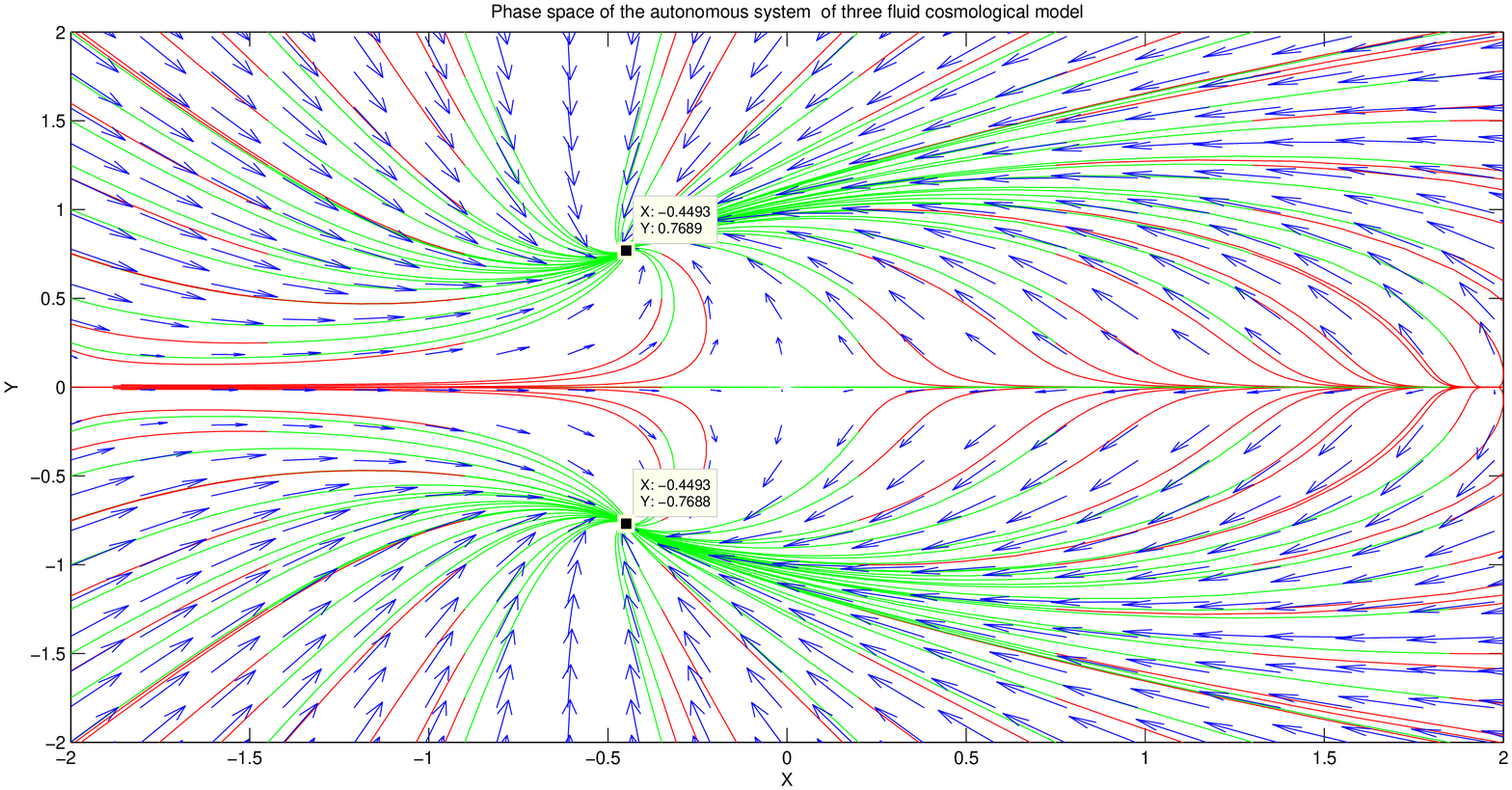}
\caption{\label{}phase portrait for the system, related table IV }
\end{minipage}
 \end{figure}

\begin{table}
 \caption{ Critical points for  $\Omega_{m} = 0.24 $, $ \nu= 1.3$ and $ \alpha= 1.3$, related  parameters  }
 \begin{tabular}{|c|c|c|c|c|c|c|}
 \hline
   Equilibrium point &  x & y  &   $  \Omega_{\phi} $ &  $\Omega_{b}$ &$  \omega_{eff} $ &  Nature \\\hline
 $ C_{2a} $  &  0  &  0 &  0 & .76 & 0.228 & saddle   \\\hline
 $ C_{7b} $  & -0.40 & 0.87 & 0.91  & .001 & 0.54 & stable node \\\hline
 $ C_{7c}$  & -0.40 & -0.87 & 0.91 & .001 & 0.54 & stable node  \\\hline

\end{tabular}
\end{table}

\begin{table}
 \caption{ Critical points for  $\Omega_{m} = 0.2$, $ \nu= 1.6$ and $ \alpha= -1.1 $, related  parameters  }
 \begin{tabular}{|c|c|c|c|c|c|c|}
 \hline
   Equilibrium point &  x & y  &   $  \Omega_{\phi} $ &  $\Omega_{b}$ &$  \omega_{eff} $ &  Nature \\\hline
 $ C_{2a} $  &  0  &  0 &  0 & 0.8 &  0.48&  saddle  \\\hline
 $ C_{4b} $  & 0.33 & 0.89 & 0.90  & .001 &-0.58 & stable node \\\hline
 $ C_{4c}$  & 0.33 & -0.89 & 0.90 & .001 &-0.58 & stable node  \\\hline

\end{tabular}
\end{table}

\begin{figure}
\begin{minipage}{.45\textwidth}
 \includegraphics[width = 1.000\linewidth]{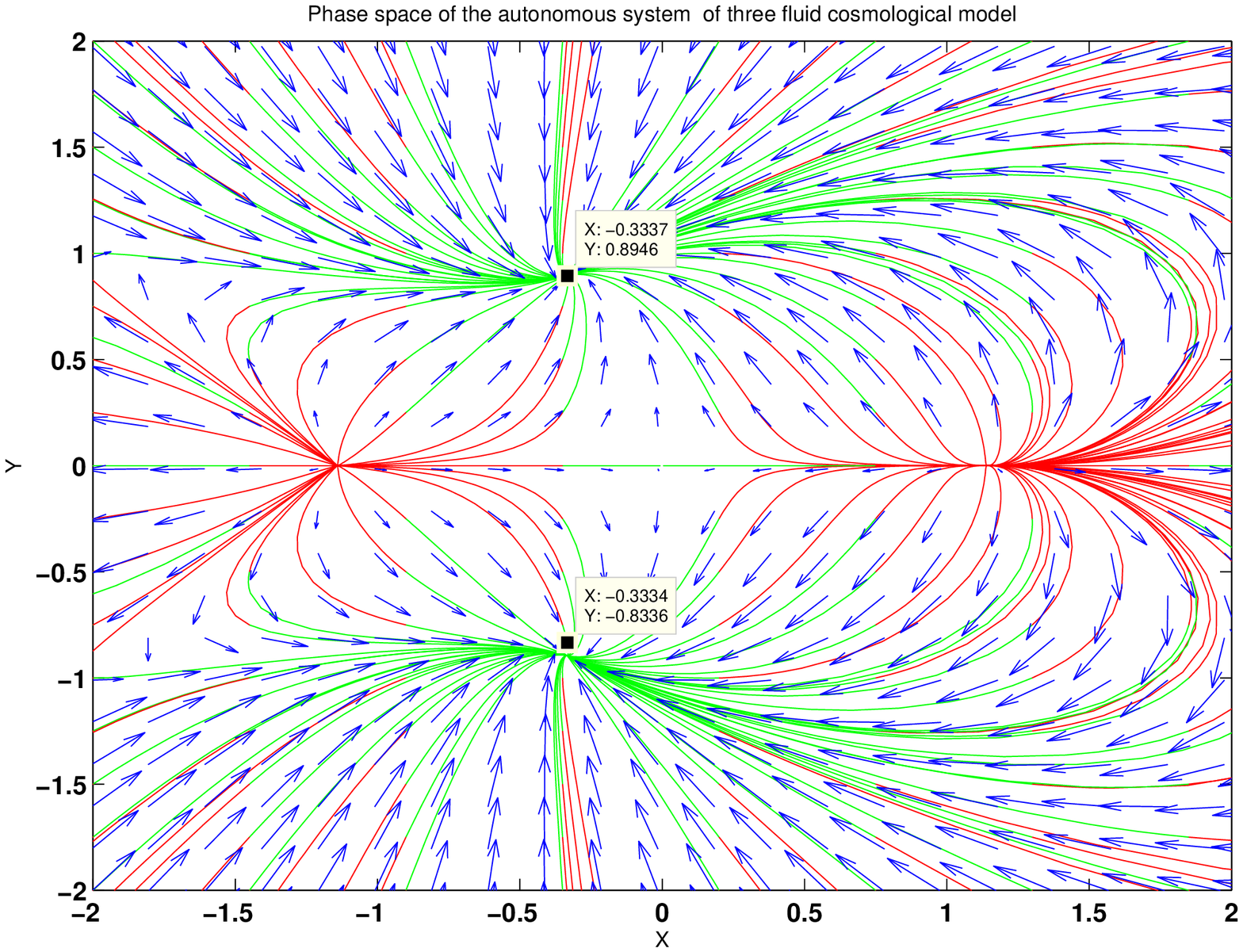}
\caption{\label{}Phase portrait for the system, related table V   }
 \end{minipage}
 \begin{minipage}{.45\textwidth}
 \includegraphics[width = 1.0\linewidth]{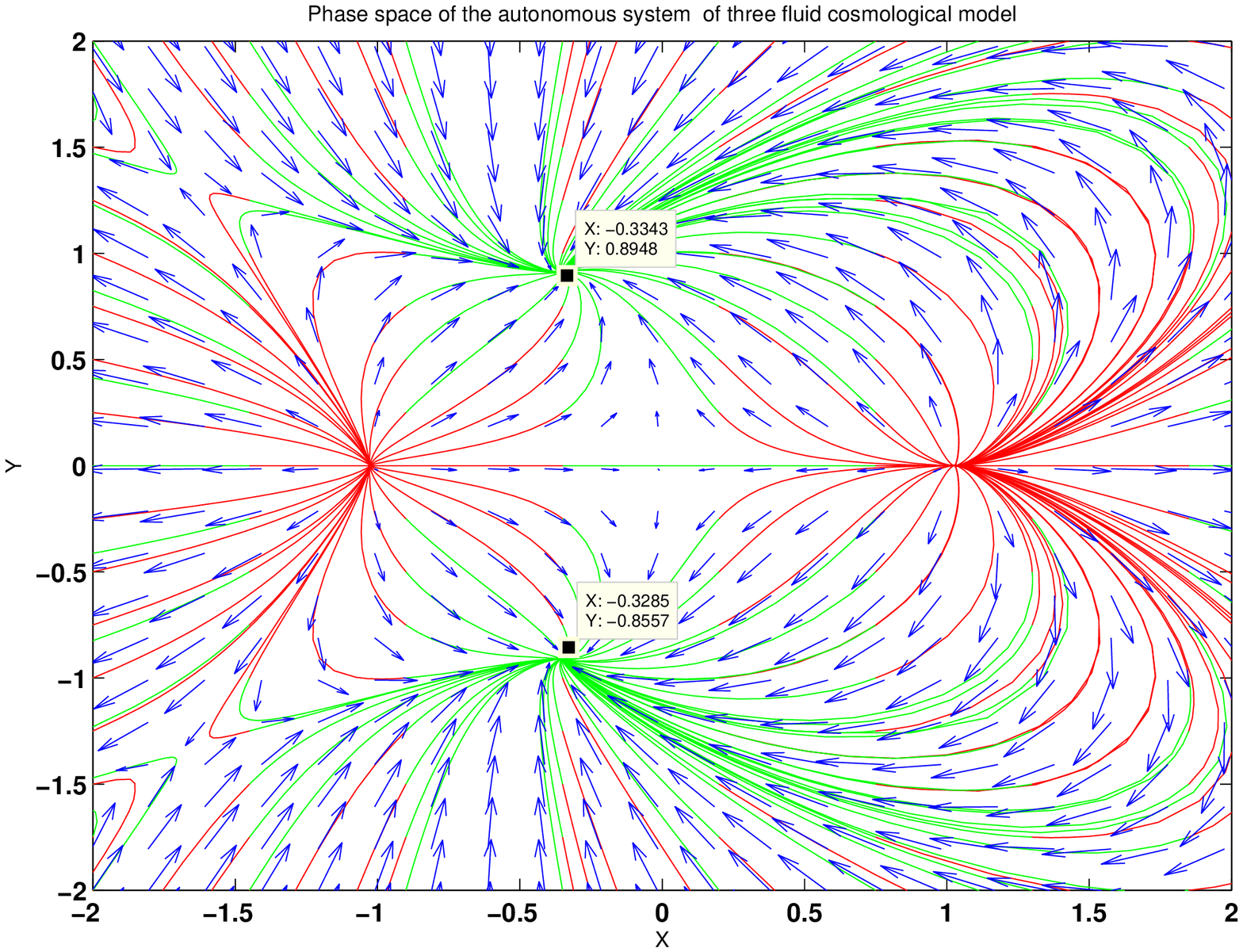}
\caption{\label{}phase portrait for the system, related table VI }
\end{minipage}
 \end{figure}

 \begin{figure}
\begin{minipage}{.45\textwidth}
 \includegraphics[width = 1.000\linewidth]{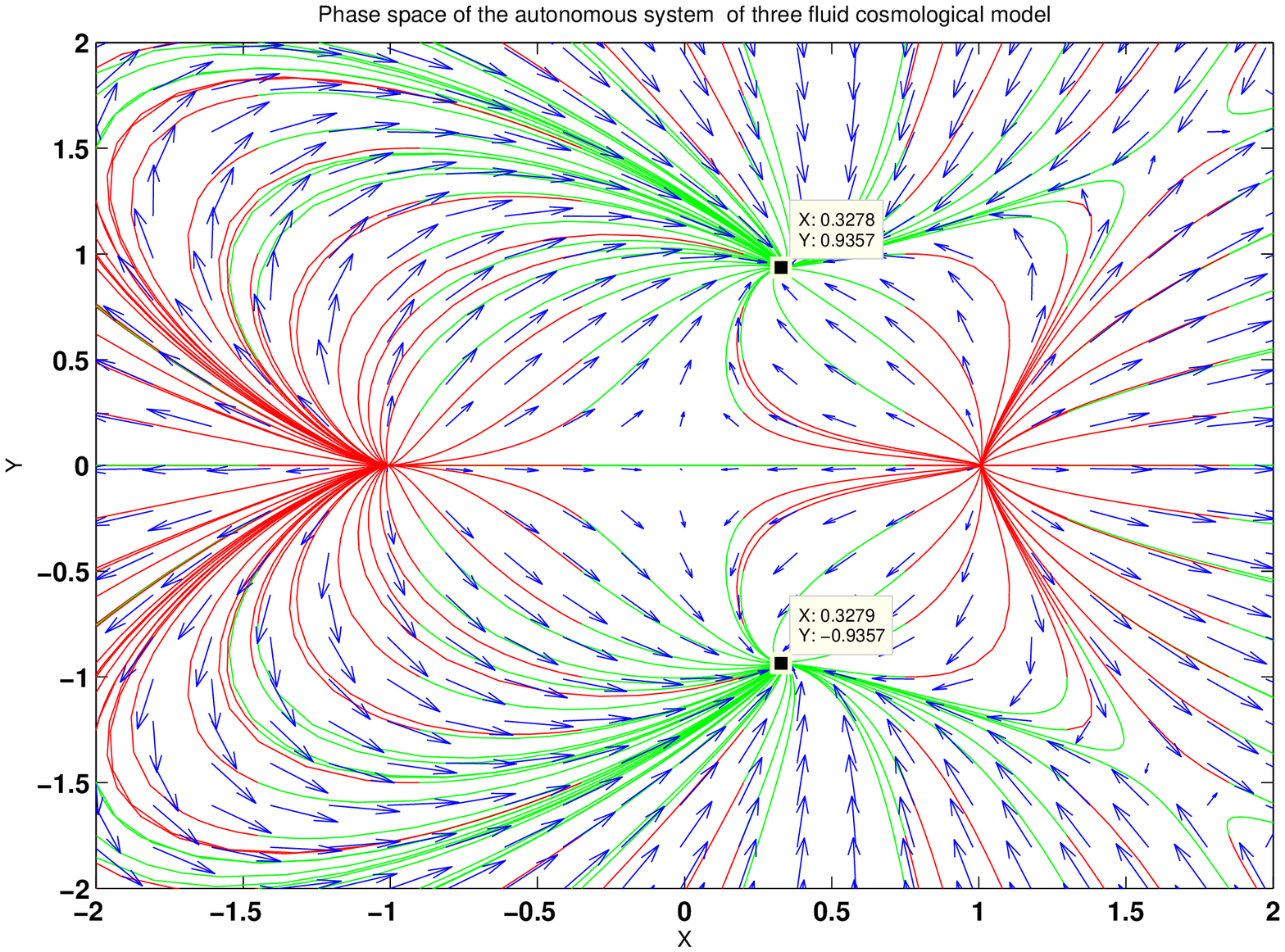}
\caption{\label{}Phase portrait for the system, related table VII     }
 \end{minipage}
 \begin{minipage}{.45\textwidth}
 \includegraphics[width = 1.00\linewidth]{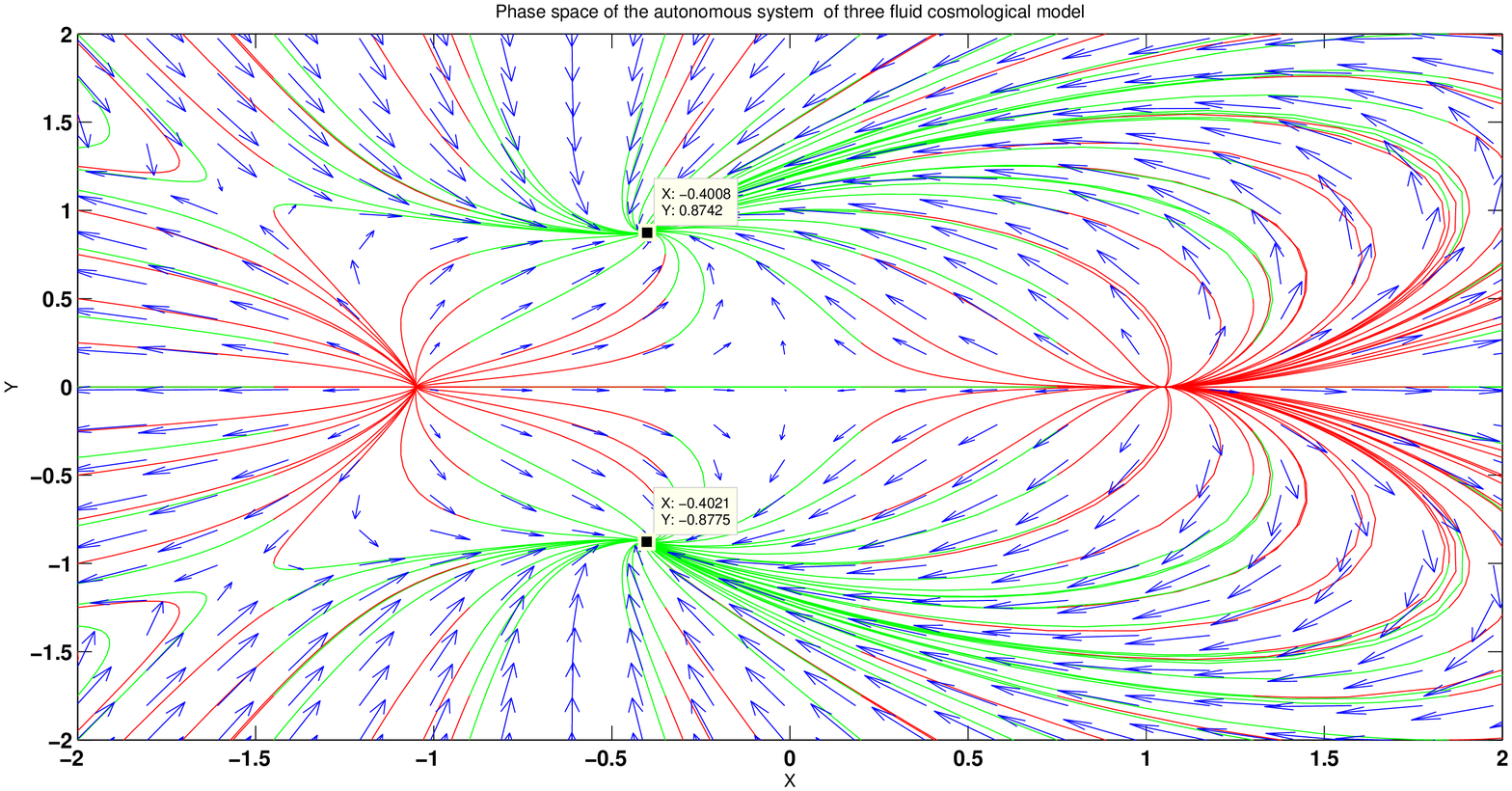}
\caption{\label{}phase portrait for the system, related table VIII }
\end{minipage}
 \end{figure}

\begin{figure}
 \includegraphics[width = .55000\linewidth]{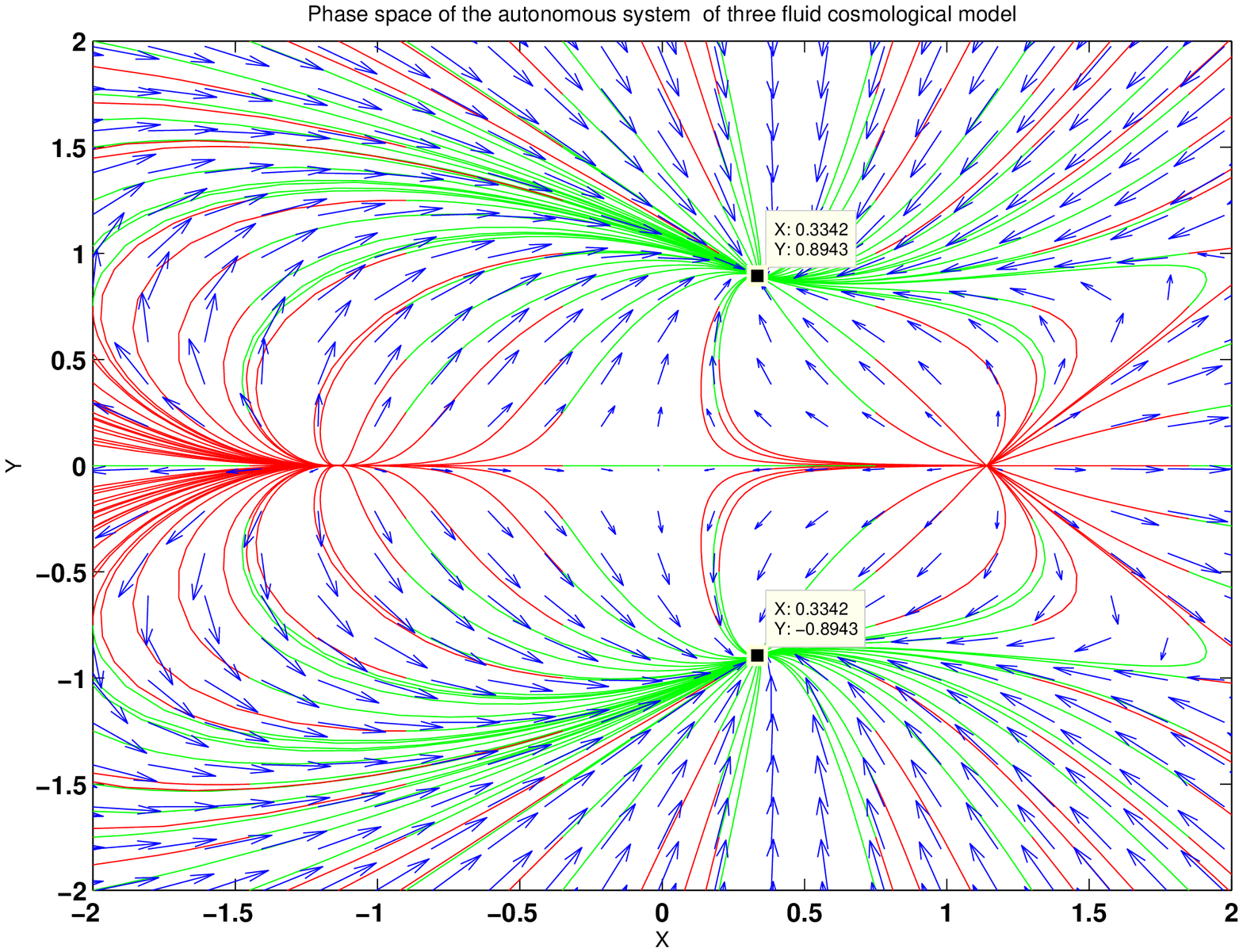}
\caption{\label{}Phase portrait for the system, related table IX }
\end{figure}

Evidently, (0, 0) is a critical point of this system. When  y = 0 and $ x \neq 0 $, we get $x = \pm \sqrt{ 1 + \frac{\Omega_{m}(\nu - 1)}{2(1-\frac{\nu}{2})}}$ and for   x = 0 and $ y \neq 0 $, we obtain $ y = \pm \sqrt{1 - \Omega_{m}+ \frac{\Omega_{m}}{\nu}}$ . We have taken the help of figures drawn for this system  for non zero x and y using MATLAB. Different critical points are obtained for different values of $\alpha$, $ \nu $. Figures are  drawn for these points. These points are listed in the  tables III to IX. From table III  (see Fig 1) and IV ( see Fig 2) we find that there are critical points, namely, $ C_{2b}, C_{2c}, C_{3b}, C_{3c}$  which are stable node in nature, but they violate (\ref{eqn(13)}) marginally.   $ C_{2d}, C_{2e}$  are  far outside of the range so are not considered. We did not list this type of critical points for other values of $\alpha$, $ \nu $ in the tables or discussion. We notice that in table V and VI ( Fig 3 and Fig 4)  $ C_{4b}, C_{4c}$  or $ C_{5b}, C_{5c}$ are stable nodes and these points  describe late time acceleration. Table VII, VIII ( Fig 5 and Fig 6 ) give critical points  $ C_{6b}, C_{6c}$  or $ C_{7b}, C_{7c}$  which are also stable node though late time acceleration is not possible for these points. Table IX ( Fig 7) gives the same critical point as in table V. All the critical points listed here like $ C_{2b}, C_{2c}$, $ C_{4b}, C_{4c}$   represent dark energy dominated universe. We need to remember $ y < 0$ represents contracting universe. This type of solution will remain invalid.
\section{ summary}
 We can see that in three dimension  no viable cosmological solution exists which satisfy both dark energy domination and late time acceleration. But in two dimension, $ C_{5b}$ is the most favorable solution representing dark energy dominated universe having late time acceleration. $  C_{5c}$ is not considered as $ y < 0$ represents contracting universe. $ C_{4b}$ may be considered another solution though bounds from  (\ref{eqn(13)}) should be remembered. Thus three fluid cosmological model may represent a cosmological solution.

\begin{acknowledgments}
 The authors are  thankful to UGC-DRS programme, Department of Mathematics, Jadavpur University.
\end{acknowledgments}
\section{References}

\end{document}